\documentclass[aps,twocolumn,showpacs,preprintnumbers,amsmath,amssymb,superscriptaddress,floatfix,nofootinbib]{revtex4}

\usepackage{graphicx}
\usepackage{epsfig}
\usepackage{epstopdf}
\usepackage{hyperref}
\usepackage{amsmath}
\usepackage{amsfonts}
\usepackage{amssymb}

\begin{document}

\title{\boldmath Role of the $f_1(1285)$ state in the $J/\psi \to \phi \bar{K} K^*$ and $J/\psi \to \phi f_1(1285)$ decays}
\date{\today}

\author{Ju-Jun Xie} \email{xiejujun@impcas.ac.cn}
\affiliation{Institute of Modern Physics, Chinese Academy of
Sciences, Lanzhou 730000, China} \affiliation{Research Center for
Hadron and CSR Physics, Institute of Modern Physics of CAS and
Lanzhou University, Lanzhou 730000, China} \affiliation{State Key
Laboratory of Theoretical Physics, Institute of Theoretical Physics,
Chinese Academy of Sciences, Beijing 100190, China}

\author{E.~Oset} \email{oset@ific.uv.es}
\affiliation{Institute of Modern Physics, Chinese Academy of
Sciences, Lanzhou 730000, China}\affiliation{Departamento de
F\'{\i}sica Te\'orica and IFIC, Centro Mixto Universidad de
Valencia-CSIC Institutos de Investigaci\'on de Paterna, Aptdo.
22085, 46071 Valencia, Spain}

\begin{abstract}

We study the role of the $f_1(1285)$ resonance in the decays of
$J/\psi \to \phi \bar{K} K^*$ and $J/\psi \to \phi f_1(1285)$. The
theoretical approach is based on the results of chiral unitary
theory where the $f_1(1285)$ resonance is dynamically generated from
the $K^* \bar{K} - c.c.$ interaction. In order to further test the
dynamical nature of the $f_1(1285)$ state, we investigate the
$J/\psi \to \phi \bar{K} K^*$ decay close to the $\bar{K} K^*$
threshold and make predictions for the ratio of the invariant mass
distributions of the $J/\psi \to \phi \bar{K} K^*$ decay and the
$J/\psi \to \phi f_1(1285)$ partial decay width with all the
parameters of the mechanism fixed in previous studies. The results
can be tested in future experiments and therefore offer new clues on
the nature of the $f_1(1285)$ state.

\end{abstract}

\maketitle

\section{Introduction}

The $f_1(1285)$ resonance [$I^G(J^{PC}) = 0^+(1^{++})$] is an
axial-vector state with mass $M_{f_1} = 1281.9 \pm 0.5$ MeV and
total decay width $\Gamma_{f_1} = 24.2 \pm 1.1$
MeV~\cite{Agashe:2014kda}. This state is described as a $q\bar{q}$
state within the quark
model~\cite{Gavillet:1982tv,Godfrey:1998pd,Li:2000dy,Vijande:2004he,Klempt:2007cp,Chen:2015iqa}.
On the other hand, the $f_1(1285)$ is also suggested to be a
dynamically generated state made from the single channel
$\bar{K}K^*$ interaction in the chiral unitary
approach~\cite{Roca:2005nm}. As shown in Ref.~\cite{Roca:2005nm},
because the $f_1(1285)$ resonance has positive $G$ parity, it cannot
couple to other pseudoscalar--vector channels. For reasons of parity
it can also not decay into two pseudoscalar mesons. Thus, since the
resonance is located below the $\bar{K} K^*$ mass threshold, its
observation is difficult in two body decays. Indeed, the main decay
channels of the $f_1(1285)$ are $4\pi$ (branching ratio $=33\%$),
$\eta \pi \pi$ ($52\%$), and $\pi \bar{K} K$ ($9\%$).

While Nature is probably more complicated and the $f_1(1285)$ state
might have components of either type (see discussions in
Ref.~\cite{Aceti:2015pma}), two comments are in order. First, the
fact that states of different nature are possible does not mean that
there should be a duplication of states with the same quantum
numbers corresponding to each type of structure. The different
structures mix and at the end it is a particular mixture what gives
rise to the observed states. These features were well described in
Refs.~\cite{vanBeveren:1986ea,Tornqvist:1995ay,Fariborz:2009cq,Fariborz:2009wf}
for the $\sigma$ ($f_0(500)$) meson. One starts with a seed of
$q\bar{q}$ and lets it couple to $\pi \pi$ components respecting
unitarity of the $\pi \pi$ interaction. At the end, a physical state
develops in which the original seed has been eaten up by the meson
cloud, which becomes the dominant component of the wave function.
The other comment is that, depending on the reaction, one or the
other component will evidence itself more clearly, and in the
present case, where we have a $\bar{K} K^*$ produced at the end, it
is quite clear that it is this component the one which will show up.

In Refs.~\cite{Aceti:2015pma,Aceti:2015zva}, the decays of
$f_1(1285) \to \eta \pi^0 \pi^0$ and $f_1(1285) \to \pi K \bar{K}$
were studied using the picture in which the $f_1(1285)$ is
dynamically generated from the single channel $\bar{K}K^*$
interaction. The theoretical predictions are compatible with the
experimental measurements. Very recently, the production of the
$f_1(1285)$ resonance in the reaction $K^- p \to f_1(1285) \Lambda$
within an effective Lagrangian approach was studied in
Ref.~\cite{Xie:2015wja} based on the results obtained in chiral
unitary theory. The theoretical calculations are in agreement with
the experimental data which provides further support for the
molecular structure of the $f_1(1285)$ state.

On the experimental side, in
Refs.~\cite{Falvard:1988fc,Jousset:1988ni}, the decay of $J/\psi \to
\phi f_1(1285)$ was studied from the $J/\psi \to \phi 2(\pi^+
\pi^-)$ and $J/\psi \to \phi \eta \pi^+ \pi^-$ decays by the DM2
Collaboration, while in Ref.~\cite{Ablikim:2007ev}, the branching
fraction of $J/\psi \to \phi \bar{K} K^*$ was measured from the
decay of $J/\psi \to \phi K \bar{K} \pi$ by the BES Collaboration.
Because the $J/\psi$ and the $\phi$ mesons have quantum numbers
$0^-(1^{--})$ and $0^-(1^{--})$, respectively, the decay $J/\psi \to
\phi \bar{K} K^*$ constitutes the ideal reaction to look for the
$f_1(1285)$ state, with quantum numbers $0^+(1^{++})$, coupling to
an $s$ wave $\bar{K} K^*$ pair. However, since the $f_1(1285)$ is
located below the $\bar{K} K^*$ threshold, it will contribute to the
region close to the threshold of $\bar{K} K^*$.

In the present work, following the formalism of
Ref.~\cite{Roca:2005nm}, we study the decays of $J/\psi \to \phi
\bar{K} K^*$ and $J/\psi \to \phi f_1(1285)$ with the picture that
the $f_1(1285)$ resonance is dynamically generated from the single
channel $\bar{K} K^*$ interaction.

This paper is organized as follows. In Sec.~\ref{sec:formalism}, we
discuss the formalism and the main ingredients of the model. In
Sec.~\ref{sec:results}, we present our main results and, finally, a
short summary and conclusions are given in Sec.~\ref{sec:summary}.

\section{Formalism} \label{sec:formalism}

We want to study the role of the $f_1(1285)$ state, which is
dynamically generated by the $\bar{K}$ and $K^*$ interaction, in the
$J/\psi \to \phi \bar{K} K^{*}$ decay. In the chiral unitary
approach of Ref.~\cite{Roca:2005nm}, the $f_1(1285)$ resonance was
obtained by solving the Bethe-Salpeter equation in the $\bar{K}K^*$
channel to obtain the scattering amplitude
\begin{eqnarray}
t = \frac{v}{1-vG},   \label{Eq:bsequation}
\end{eqnarray}
where $v$ is the $\bar{K}K^* \to \bar{K}K^*$ transition potential
and $G$ is the loop function for the propagators of the $\bar{K}$
and $K^*$ mesons given in Ref.~\cite{Roca:2005nm}. The $v$ and $G$
depend on the invariant mass $M_{\rm inv}$ of the $\bar{K}K^*$
system, and hence the scattering amplitude $t$ is also dependent on
$M_{\rm inv}$. The loop function $G$ is divergent, and it can be
regularized both with a cutoff prescription or with dimensional
regularization in terms of a subtraction
constant~\cite{Oller:2000fj}. In this work we will make use of the
cutoff regularization scheme, which introduces a cutoff parameter
$q_{\rm max}$. The cut off is tuned to get a pole of the $t$ matrix
at the mass (1281.3 MeV) of the $f_1(1285)$. This provides the
coupling $g_{f_1} = 7555$ MeV of the resonance to the $\bar{K}K^*$
channel (see more details in Ref.~\cite{Aceti:2015pma}). With the
explicit expressions for $v$ and $G$ taken from
Ref.~\cite{Roca:2005nm}, we obtain a good description of the
$f_1(1285)$ resonance using a cutoff $q_{\rm max} = 990$ MeV, as in
Ref.~\cite{Roca:2005nm}.

\begin{figure*}[htbp]
\begin{center}
\includegraphics[scale=0.8]{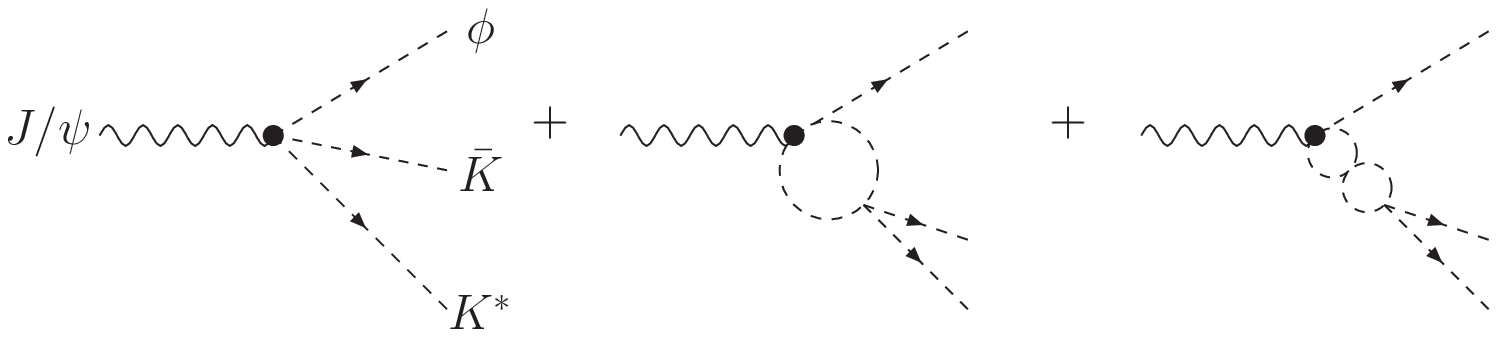} \vspace{1.cm}
\caption{Diagrammatic representation of the $J/\psi \to \phi \bar{K}
K^{*}$ decay. \label{Fig:jpsitophikbarkstar}}
\end{center}
\end{figure*}

For $J/\psi \to \phi \bar{K} K^{*}$, the decay mechanism is shown in
Fig.~\ref{Fig:jpsitophikbarkstar}. To take into account the final
state interaction of the $\bar{K} K^*$ pair, we have to consider the
resummation of the diagrams shown in the figure.

According to the diagrams in Fig.~\ref{Fig:jpsitophikbarkstar}, the
transition matrix for the process $J/\psi \to \phi \bar{K} K^*$ can
be given by
\begin{eqnarray}
T_{J/\psi \to \phi \bar{K}K^*} &=& V_P C_s \left[1 +
G(M_\text{inv}^2)
t(M_\text{inv}^2) \right] \nonumber \\
&=& V_P C_s \frac{t(M_\text{inv})}{v(M_\text{inv})}~,
\label{eq:amplitudephiKKstar}
\end{eqnarray}
where the last equality follows from Eq.~\eqref{Eq:bsequation}. The
$V_P$ and $C_s$ are the bare production vertex and the spin
structure (the spin of $K^*$ together with the one of the $\phi$
must give the spin of $J/\psi$: $1 \bigotimes 1 \to 1$) factor for
$J/\psi \to \phi \bar{K} K^{*}$. We assume that this bare vertex is
of a short range nature, {\it i.e.}, just a coupling constant in the
field theory language.

The spin structure of the $J/\psi$, $K^*$, and $\phi$ coupling can
be written as
\begin{eqnarray}
C_s = \epsilon_{ijk} \varepsilon_i(J/\psi) \varepsilon_j(\phi)
\varepsilon_k(K^*) .
\end{eqnarray}

Summing and averaging $C^2_s$ over final and initial polarizations
of the vector mesons we find
\begin{eqnarray}
\overline{\sum} \sum C^2_s &=& \frac{2}{3} (3 +
\frac{p^2_{\phi}}{m^2_{\phi}} + \frac{p^2_{K^*}}{m^2_{K^*}}),
\end{eqnarray}
where $p_{\phi}$ and and $p_{K^*}$ are the $\phi$ and $K^*$ momenta
in the $J/\psi$ rest frame, respectively,
\begin{eqnarray}
p_{\phi} &=& \frac{\lambda^{1/2}(M^2_{J/\psi}, m^2_{\phi},
M^2_\text{inv})}{2M_{J/\psi}}, \\
p_{K^*} &=& \frac{\lambda^{1/2}(M^2_{J/\psi}, m^2_{K^*}, M^2_{\phi
\bar{K}})}{2M_{J/\psi}},
\end{eqnarray}
where $M_{\phi \bar{K}}$ is the invariant mass of $\phi \bar{K}$
system, and $\lambda(x,y,z)$ is the K\"ahlen or triangle function.

We can easily get the $\bar{K} K^{*}$ invariant mass spectrum for
the $J/\psi \to \phi \bar{K} K^{*}$
as~\cite{Nacher:1998mi,MartinezTorres:2012du,Xie:2013ula}:
\begin{eqnarray}
\frac{d\Gamma_{J/\psi \to \phi \bar{K}K^*}}{d M_{\text{inv}}} &=&
\frac{V^2_p}{(2\pi)^3} \frac{M_{\rm inv}}{8 M^3_{J/\psi}}
\left\lvert
\frac{t(M_\text{inv})}{v(M_{\rm inv})} \right\rvert^2 \nonumber \\
&& \times \int^{M^{\rm max}_{\phi \bar{K}}}_{M^{\rm min}_{\phi
\bar{K}}} \overline{\sum} \sum C^2_s M_{\phi \bar{K}} dM_{\phi
\bar{K}}. \label{eq:dgdm}
\end{eqnarray}
For a given value of $M_{\rm inv}$, the range of $M_{\phi \bar{K}}$
is defined as,
\begin{eqnarray}
M^{\rm max}_{\phi \bar{K}} \! &=& \! \sqrt{ \left(E_{\bar{K}} +
E_{\phi} \right)^2-\left(\sqrt{E^{2}_{\bar{K}} - m^2_{\bar{K}}} - \sqrt{E^{2}_{\phi} - m^2_{\phi}} \right)^2 }, \nonumber \\
M^{\rm min}_{\phi \bar{K}} \! &=& \! \sqrt{ \left(E_{\bar{K}} +
E_{\phi} \right)^2 - \left(\sqrt{E^{2}_{\bar{K}} - m^2_{\bar{K}}} +
\sqrt{E^{2}_{\phi} - m^2_{\phi}} \right)^2 }, \nonumber
\end{eqnarray}
where $E_{\bar{K}} = (M^2_{\rm inv} - m^2_{K^*} +
m^2_{\bar{K}})/2M_{\rm inv}$ and $E_{\phi} = (M^2_{J/\psi} -
M^2_{\rm inv} - m^2_{\phi})/2M_{\rm inv}$ are the energies of
$\bar{K}$ and $\phi$ in the $\bar{K} K^*$ rest frame.

\begin{figure}[htbp]
\begin{center}
\includegraphics[scale=0.9]{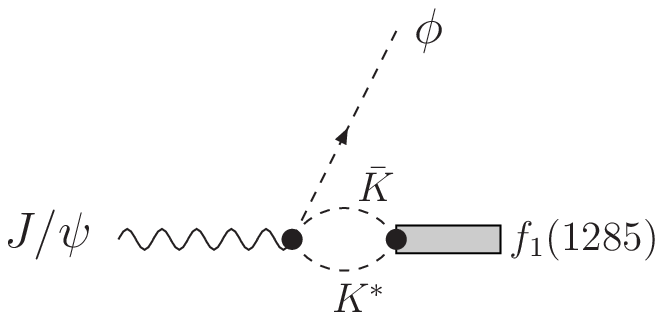}
\caption{Production mechanism of the $J/\psi \to \phi f_1(1285)$
decay. \label{Fig:jpsitophif1}}
\end{center}
\end{figure}

On the other hand, if we are interested in the production of the
$f_1(1285)$ resonance, the relevant mechanism is depicted
diagrammatically in Fig.~\ref{Fig:jpsitophif1} and we have
\begin{eqnarray}
T_{J/\psi \to \phi f_1(1285)} = V_P C'_s G(M_{f_1}) g_{f_1},
\end{eqnarray}
where the spin factor $C'_s$ is easily obtained. We must recall that
the coupling of $f_1(1285)$ to $\bar{K}K^* - c.c.$ is given by
$g_{f_1} \varepsilon_i(f_1)  \varepsilon_i(K^*)$. Contracting the
two $\varepsilon(K^*)$ in the $K^*$ propagator in
Fig.~\ref{Fig:jpsitophif1} we have
\begin{eqnarray}
C'_s = \epsilon_{ijk} \varepsilon_i(J/\psi) \varepsilon_j(\phi)
\varepsilon_k(f_1).
\end{eqnarray}

Then, the partial decay width of $J/\psi \to \phi f_1(1285)$ is
given by
\begin{eqnarray}
\Gamma_{J/\psi \to \phi f_1(1285)} = \frac{V^2_P}{8\pi}
\frac{G^2(M_{f_1}) g^2_{f_1} p'_{\phi}}{M^2_{J/\psi}}
\overline{\sum} \sum {C'_s}^2, \label{eq:gamrf1}
\end{eqnarray}
with
\begin{eqnarray}
\overline{\sum} \sum {C'_s}^2 = \frac{2}{3} (3 +
\frac{{p'}^2_{\phi}}{M^2_{f_1}} + \frac{{p'}^2_{\phi}}{m^2_{\phi}}),
\end{eqnarray}
and $p'_{\phi}$ is the $\phi$ meson momentum obtained in the
$J/\psi$ rest frame which is
\begin{eqnarray}
p'_{\phi} &=& \frac{\lambda^{1/2}(M^2_{J/\psi}, m^2_{\phi},
M^2_{f_1})}{2M_{J/\psi}}.
\end{eqnarray}

The chiral theory cannot provide the value of the constant $V_P$ in
Eqs.~\eqref{eq:dgdm} and \eqref{eq:gamrf1}, however, if we divide
$d\Gamma/dM_{\rm inv}$ by $\Gamma_{J/\psi \to \phi f_1(1285)}$ the
constant $V_P$ is cancelled, and we can make precise predictions for
the ratio $R_{\Gamma}$ as,
\begin{eqnarray}
R_{\Gamma} = \frac{d\Gamma_{J/\psi \to \phi \bar{K}K^*}/dM_{\rm
inv}}{\Gamma_{J/\psi \to \phi f_1(1285)}}.  \label{eq:ratio}
\end{eqnarray}
This ratio is relevant because it has no free parameters (all the
parameters are fixed by previous works) and, thus, it is a
prediction of the theory. The shape, as well as the absolute values
of the ratio $R_{\Gamma}$ for the $\bar{K}K^*$ mass distribution,
can be compared with the experimental measurements.

\section{Numerical results and discussion} \label{sec:results}

\begin{figure}[htbp]
\begin{center}
\includegraphics[scale=0.45]{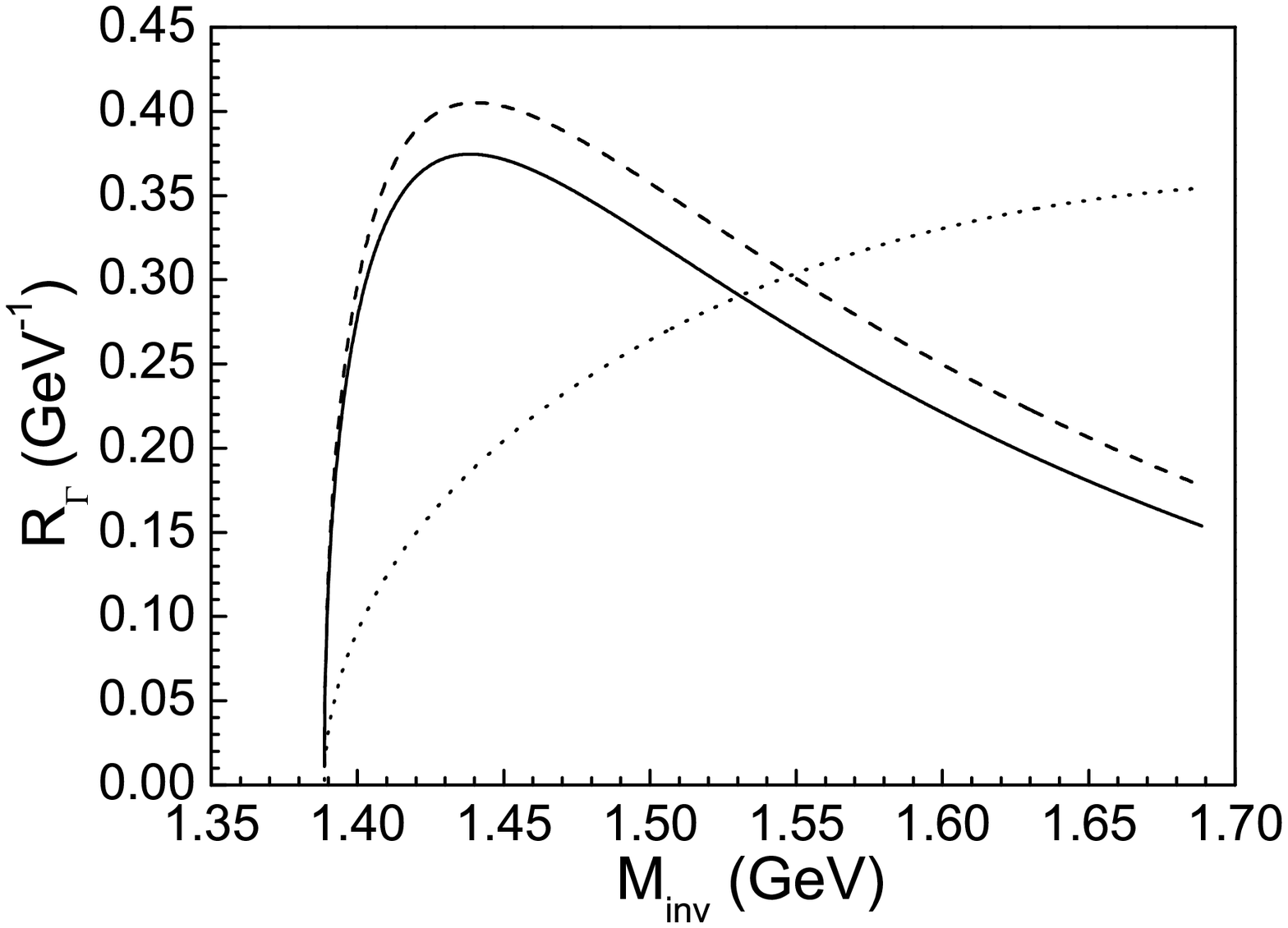}
\caption{Results of $R_{\Gamma}$ as a function of invariant mass
$M_{\rm inv}$ of $\bar{K}K^{*}$. The solid (with the spin structure
factor) and dashed (without the spin structure factor) curves stand
for the theory predictions and the dotted curve stands for the phase
space. The dotted curve is normalized such as to have the same area
as the solid curve in the $M_{\rm inv}$ range of the figure.
\label{fig:ratio}}
\end{center}
\end{figure}

In Fig.~\ref{fig:ratio}, the numerical results of $R_{\Gamma}$ as a
function of the invariant mass $M_{\rm inv}$ of the $\bar{K}K^{*}$
system are shown. The solid curve stands for the theory prediction
and the dotted curve stands for the phase space. For evaluating the
contributions of the phase space, we replace $t(M_{\rm
inv})/v(M_{\rm inv})$ of Eq.~\eqref{eq:amplitudephiKKstar} by a
constant, thus removing any effect of the $M_{\rm inv}$ dependence
of the $f_1(1285)$ resonance. Then we tune this constant such that
the $M_{\rm inv}$ integrated $R_{\Gamma}$ in the range of energies
from the $\bar{K}K^*$ threshold to $1.7$ GeV is the same as the one
evaluated with the explicit resonance formalism.

In addition, in Fig.~\ref{fig:ratio} we also show the results which
are obtained without considering the spin structure factor by the
dashed curve in Fig.~\ref{fig:ratio}. We see that the structure
factor gives a small effect to our predictions and could be
neglected.

We see a clear threshold enhancement in Fig.~\ref{fig:ratio} which
is caused by the contributions of the $f_1(1285)$ state below
threshold, which is dynamically generated by the $\bar{K}K^*$
interaction. The theoretical predictions can be tested by future
experiments.

Actually, the range of the invariant mass of $\bar{K}K^*$ in the
decay of $J/\psi \to \phi \bar{K} K^*$ is from the threshold of
$\bar{K}K^*$ up to $2.077$ GeV ($M_{J/\psi} - m_{\phi} = 2077$ MeV),
however, we cannot go so far because the chiral theory works well
about $200 - 300$ MeV from the threshold, hence we consider only the
range of $300$ MeV above the $\bar{K}K^*$ threshold as shown in
Fig.~\ref{fig:ratio}.

On the other hand, experimentally we have, from
Ref.~\cite{Agashe:2014kda},
\begin{eqnarray}
Br(J/\psi \to \phi \bar{K} K^*) &=& (2.18 \pm 0.23) \times 10^{-3}, \label{eq:brjpsitophikbarkstar}\\
Br(J/\psi \to \phi f_1(1285)) &=& (6.00 \pm 3.16) \times 10^{-4}.
\label{eq:brjpsitophif1}
\end{eqnarray}

Note that we have corrected the branching ratio $Br(J/\psi \to \phi
f_1(1285)) = (2.6 \pm 0.5) \times 10^{-4}$ quoted in the PDG which
we found is misquoted. In Ref.~\cite{Jousset:1988ni}, from where the
PDG information is obtained, the peak around $1297$ MeV of the $\eta
\pi^+ \pi^-$ mass distribution was attributed to the $f_1(1285)$
with a width of $10 \pm 8$ MeV. They obtain $Br(J/\psi \to \phi
X(1297)) \times Br(X(1297) \to \eta \pi^+ \pi^-) = (2.1 \pm 0.5 \pm
0.4) \times 10^{-4}$. Taking now into account that $Br(f_1(1285) \to
\eta \pi^+ \pi^-) = (35 \pm 15)\%$ from the PDG, we obtain
$Br(J/\psi \to \phi f_1(1285)) = (6.00 \pm 3.16) \times 10^{-4}$ as
shown in Eq.~\eqref{eq:brjpsitophif1}.

Note that we do not use the value of $Br(J/\psi \to \phi f_1(1285))
= (3.2 \pm 0.6 \pm 0.4) \times 10^{-4}$ presented in
Ref.~\cite{Jousset:1988ni}, which was obtained from the decay of
$J/\psi \to \phi 2(\pi^+ \pi^-)$. Very recently, in
Ref.~\cite{Ablikim:2014pfc}, the branching fraction $Br(J/\psi \to
\phi f_1(1285), f_1(1285) \to \eta \pi^+ \pi^-) $ was measured, with
the result $(1.2 \pm 0.06 \pm 0.14) \times 10^{-4}$. Taking this
value into account, and the $Br(f_1(1285) \to \eta \pi^+ \pi^-)$
used before, we get $Br(J/\psi \to \phi f_1(1285)) = (3.43 \pm 1.53)
\times 10^{-4}$. This value is consistent within errors with what we
have obtained before.

From Eqs.~\eqref{eq:brjpsitophikbarkstar} and
\eqref{eq:brjpsitophif1} we obtain
\begin{eqnarray}
R = \frac{Br(J/\psi \to \phi \bar{K} K^*)}{Br(J/\psi \to \phi
f_1(1285))} = 3.6 \pm 2.0 ~ .  \label{eq:rexp}
\end{eqnarray}
One might think we should compare our theoretical result, $R =
\int^{M_{J/\psi} - m_{\phi}}_{m_{\bar{K}} + m_{K^*}} R_{\Gamma}
dM_{\rm inv}$ , to the experimental result in Eq.~\eqref{eq:rexp},
but, as discussed before, we take the $\bar{K}K^* \to \bar{K}K^*$
scattering amplitude $t(M_{\rm inv})$ from the chiral unitary
approach, and we can not go too far from the $\bar{K}K^*$ threshold.
Furthermore, there could be also other contributions from higher
mass states with spin-parity $J^P = 1^+$ and $2^+$ at higher
invariant mass region of $\bar{K}K^*$. These higher states will not
contribute too much to the lower energy region and hence will not
affect our predictions here. On the other hand, note that the
experimental results of Ref.~\cite{Jousset:1988ni} were obtained in
the 1980s and only few signal events were observed. Further
improvement can be done in the future at BESIII or BelleII. The
future experimental observation of the mass distribution
$R_{\Gamma}$ would provide very valuable information on the
mechanism of the $J/\psi \to \phi \bar{K}K^*$ decay.

\section{Summary} \label{sec:summary}

In summary, we have studied the decays of $J/\psi \to \phi \bar{K}
K^{*}$ and $J/\psi \to \phi f_1(1285)$ with the theoretical approach
which is based on results of chiral unitary theory where the
$f_1(1285)$ resonance is dynamically generated from the $K^* \bar{K}
- c.c.$ interaction. The ratio $R_{\Gamma} = \frac{d\Gamma_{J/\psi
\to \phi \bar{K}K^*}/dM_{\rm inv}}{\Gamma_{J/\psi \to \phi
f_1(1285)}}$ as a function of invariant mass $M_{\rm inv}$ of
$\bar{K}K^{*}$ is predicted. A clear threshold enhancement in
Fig.~\ref{fig:ratio} compared with the phase space appears, which is
caused by the presence of the $f_1(1285)$ state below threshold .
The experimental observation of this mass distribution would then
provide very valuable information to check our predictions and the
basic nature of the $f_1(1285)$ resonance.

\section*{Acknowledgments}

One of us, E. O., wishes to acknowledge support from the Chinese
Academy of Science in the Program of Visiting Professorship for
Senior International Scientists (Grant No. 2013T2J0012). This work
is partly supported by the Spanish Ministerio de Economia y
Competitividad and European FEDER funds under the contract number
FIS2011-28853-C02-01 and FIS2011-28853-C02-02, and the Generalitat
Valenciana in the program Prometeo II-2014/068. We acknowledge the
support of the European Community-Research Infrastructure
Integrating Activity Study of Strongly Interacting Matter (acronym
HadronPhysics3, Grant Agreement n. 283286) under the Seventh
Framework Programme of EU. This work is also partly supported by the
National Natural Science Foundation of China under Grant No.
11475227. This work is also supported by the Open Project Program of
State Key Laboratory of Theoretical Physics, Institute of
Theoretical Physics, Chinese Academy of Sciences, China
(No.Y5KF151CJ1).

\bibliographystyle{plain}

\begin{thebibliography}{999}


\bibitem{Agashe:2014kda}
  K.~A.~Olive {\it et al.}  [Particle Data Group Collaboration],
  Chin.\ Phys.\ C {\bf 38}, 090001 (2014).

\bibitem{Gavillet:1982tv}
  P.~Gavillet, R.~Armenteros, M.~Aguilar-Benitez, M.~Mazzucato and C.~Dionisi,
  Z.\ Phys.\ C {\bf 16}, 119 (1982).

\bibitem{Godfrey:1998pd}
  S.~Godfrey and J.~Napolitano,
  Rev.\ Mod.\ Phys.\  {\bf 71}, 1411 (1999).

\bibitem{Li:2000dy}
  D.~M.~Li, H.~Yu and Q.~X.~Shen,
  Chin.\ Phys.\ Lett.\  {\bf 17}, 558 (2000).

\bibitem{Vijande:2004he}
  J.~Vijande, F.~Fernandez and A.~Valcarce,
  J.\ Phys.\ G {\bf 31}, 481 (2005).


\bibitem{Klempt:2007cp}
  E.~Klempt and A.~Zaitsev,
  Phys.\ Rept.\  {\bf 454}, 1 (2007).


\bibitem{Chen:2015iqa}
  K.~Chen, C.~Q.~Pang, X.~Liu and T.~Matsuki,
  Phys.\ Rev.\ D {\bf 91}, 074025 (2015).


\bibitem{Roca:2005nm}
  L.~Roca, E.~Oset and J.~Singh,
  Phys.\ Rev.\ D {\bf 72}, 014002 (2005).

\bibitem{Aceti:2015pma}
  F.~Aceti, J.~J.~Xie and E.~Oset,
  Phys.\ Lett.\ B {\bf 750}, 609 (2015).

\bibitem{vanBeveren:1986ea}
  E.~van Beveren, T.~A.~Rijken, K.~Metzger, C.~Dullemond, G.~Rupp and J.~E.~Ribeiro,
  Z.\ Phys.\ C {\bf 30}, 615 (1986).

\bibitem{Tornqvist:1995ay}
  N.~A.~Tornqvist and M.~Roos,
  Phys.\ Rev.\ Lett.\  {\bf 76}, 1575 (1996).

\bibitem{Fariborz:2009cq}
  A.~H.~Fariborz, R.~Jora and J.~Schechter,
  Phys.\ Rev.\ D {\bf 79}, 074014 (2009).

\bibitem{Fariborz:2009wf}
  A.~H.~Fariborz, N.~W.~Park, J.~Schechter and M.~Naeem Shahid,
  Phys.\ Rev.\ D {\bf 80}, 113001 (2009).

\bibitem{Aceti:2015zva}
  F.~Aceti, J.~M.~Dias and E.~Oset,
  Eur.\ Phys.\ J.\ A {\bf 51}, 48 (2015).

\bibitem{Xie:2015wja}
  J.~J.~Xie,
  Phys.\ Rev.\ C {\bf 92}, 065203 (2015).

\bibitem{Falvard:1988fc}
  A.~Falvard {\it et al.} [DM2 Collaboration],
  Phys.\ Rev.\ D {\bf 38}, 2706 (1988).

\bibitem{Jousset:1988ni}
  J.~Jousset {\it et al.} [DM2 Collaboration],
  Phys.\ Rev.\ D {\bf 41}, 1389 (1990).

\bibitem{Ablikim:2007ev}
  M.~Ablikim {\it et al.} [BES Collaboration],
  Phys.\ Rev.\ D {\bf 77}, 032005 (2008).

\bibitem{Oller:2000fj}
  J.~A.~Oller and U.~G.~Meissner,
  Phys.\ Lett.\ B {\bf 500}, 263 (2001).

\bibitem{Nacher:1998mi}
  J.~C.~Nacher, E.~Oset, H.~Toki and A.~Ramos,
  Phys.\ Lett.\ B {\bf 455}, 55 (1999).

\bibitem{MartinezTorres:2012du}
  A.~Martinez Torres, K.~P.~Khemchandani, F.~S.~Navarra, M.~Nielsen and E.~Oset,
  Phys.\ Lett.\ B {\bf 719}, 388 (2013).


\bibitem{Xie:2013ula}
  J.~J.~Xie, M.~Albaladejo and E.~Oset,
  Phys.\ Lett.\ B {\bf 728}, 319 (2014).

\bibitem{Ablikim:2014pfc}
   M.~Ablikim {\it et al.} [BESIII Collaboration],
   Phys.\ Rev.\ D {\bf 91}, 052017 (2015).

\end{thebibliography}

\end{document}